\documentclass[aps,prl,twocolumn,superscriptaddress]{revtex4-1}

\usepackage{graphicx}
\usepackage{amsmath}
\usepackage{color}
\usepackage{gensymb}
\usepackage[caption=false]{subfig}

\begin{document}




\title{Phonon-mediated High-voltage Detector with Background Rejection for Low-mass Dark Matter and Reactor Coherent Neutrino Scattering Experiments}


\author{H. Neog}
\author{R. Mahapatra}
\author{N. Mirabolfathi}
\author{M. Platt}
\author{A. Jastram}
\author{G. Agnolet}
\author{A. Kubik}
\author{H. Chen}
\affiliation{Department of Physics and Astronomy, Texas A\&M University, College Station, TX, 77843}
\author{B. Mohanty}
\affiliation{School of Physical Sciences, National Institute of Science Education and Research, HBNI, Jatni - 752050, India}    

\date{\today}

\begin{abstract}
We report the first demonstration of a phonon-mediated silicon detector technology that provides a primary phonon measurement in a low-voltage region, and a simultaneous indirect measurement of the ionization signal through Neganov-Trofimov-Luke amplification in a high voltage region, both in a monolithic crystal. We present characterization of charge and phonon transport between the two stages of the detector and the resulting background discrimination capability at low energies. This new detector technology has the potential to significantly enhance the sensitivity of dark matter and coherent neutrino scattering experiments beyond the capabilities of current technologies that have limited discrimination at low energies.

\end{abstract}


\maketitle

Numerous astronomical observations point to the presence of a non-luminous, non-baryonic form of matter (called Dark Matter) with an abundance of about five times the baryonic matter in the Universe. Although there is no consensus about its composition, Weakly Interacting Massive Particles (WIMPs) form a leading candidate for the Dark Matter with interaction strengths in the weak interaction scale and a broad range of masses
~\cite{Jungman1996,Lin2017,Goodman1985}. A number of terrestrial direct detection experiments study the elastic scattering of WIMPs off nuclei in well shielded underground detectors~\cite{Bernabei2010,Aalseth2011,Angloher2012,Agnese2013,edelweiss2014,xenon102011,xenon1t2019,lux2017}. Such WIMP recoils from nuclei lead to a featureless exponentially decreasing
distribution of deposited energy that 
necessitates low-threshold detector technologies to maximize WIMP detection sensitivity~\cite{Jungman1996,Lewin1996}. 
Energy deposition in such detectors from environmental backgrounds also tend to rise sharply at low energies, thus making the search for WIMPs a very challenging effort, especially for low-mass WIMPs. 

Many experiments use techniques based on simultaneous measurement of multiple phenomena (ionization, heat, light) associated with particle interactions in order to discriminate the signal nuclear recoil (NR) from the electron recoil (ER) that is expected from dominant radioactive backgrounds~\cite{Angloher2012, Agnese:2017jvy, xenon102011}. Due to the signal to noise limitations, preserving the discrimination down to very low energy recoils is very challenging. Thus, these experiments generally abandon their background rejection capabilities by measuring only one form of amplified signal to achieve low recoil energy thresholds and thus low-mass WIMP sensitivity~\cite{Agnese2014,xenon102011}. The detector technology being reported in this letter demonstrates event-by-event discrimination between ER and NR events at low energies through phonon only measurements in a unique new detector design. 
This new detector technology holds the potential to significantly enhance the sensitivity for low-mass WIMP searches as well as searches for new physics in Coherent Elastic Neutrino Nucleus Scattering (CE$\nu$NS) experiments~\cite{Agnese:2016cpb,Agnolet:2016zir}.  

The Cyrogenic Dark Matter Search (CDMS) Z-sensitive Ionization and Phonon (ZIP) detector technology was designed to provide event-by-event discrimination between NR and ER events through simultaneous measurement of the ionization signal and the phonon signal~\cite{Ahmed:2009zw}. The primary phonons produced during the recoil don't depend on the type of particle interaction, thus providing a true measure of the recoil energy. On the other hand, the ionization produced during the recoil strongly depends on the type of particle interaction. 
Due to the fundamental differences in ionization processes in semiconductors, NR interactions are substantially less ionizing compared to the ER interactions of the equivalent recoil energy~\cite{Lindhard1961}. The Super Cryogenic Dark Matter Search (SuperCDMS) also uses this discrimination technique in the interleaved-ZIP (iZIP) detectors, by incorporating the ionization and phonon sensors in an interleaved fashion to provide further rejection of surface events~\cite{Agnese2013_2}.


The ionization to phonon energy ratio has been used as the discriminator in CDMS and the SuperCDMS experiments~\cite{Ahmed:2009zw}. For the ionization measurement, due to the inherent high impedance readout noise that depends on the equivalent capacitance of the detector and the readout front-end amplifiers, this discrimination ceases to be effective for energies below $\sim$1 keV. The ionization readout noise limitation can be solved by indirect detection of ionization via the measurement of the phonons released during the charge carriers' drift in the bulk of the Ge or Si crystals. This Neganov-Trofimov-Luke (NTL) phonon energy gain~\cite{Neganov1993} scales linearly with the bias voltage while the phonon readout noise stays the same, up to the level where leakage current becomes the dominant noise. SuperCDMS high voltage (HV) detectors use this very low noise technology in order to detect recoils down to 100 eV and to achieve world-leading sensitivity to low-mass WIMPs ($< 10 GeV/c^2$)~\cite{Agnese:2017jvy}. However, this very low threshold detection is achieved at the expense of the NR/ER discrimination because in this method the phonon readout is effectively proportional to the ionization, thus overshadowing the information from the recoil phonons. 

\begin{figure}[htp]
	
	\subfloat[]{%
		\includegraphics[width=\linewidth]{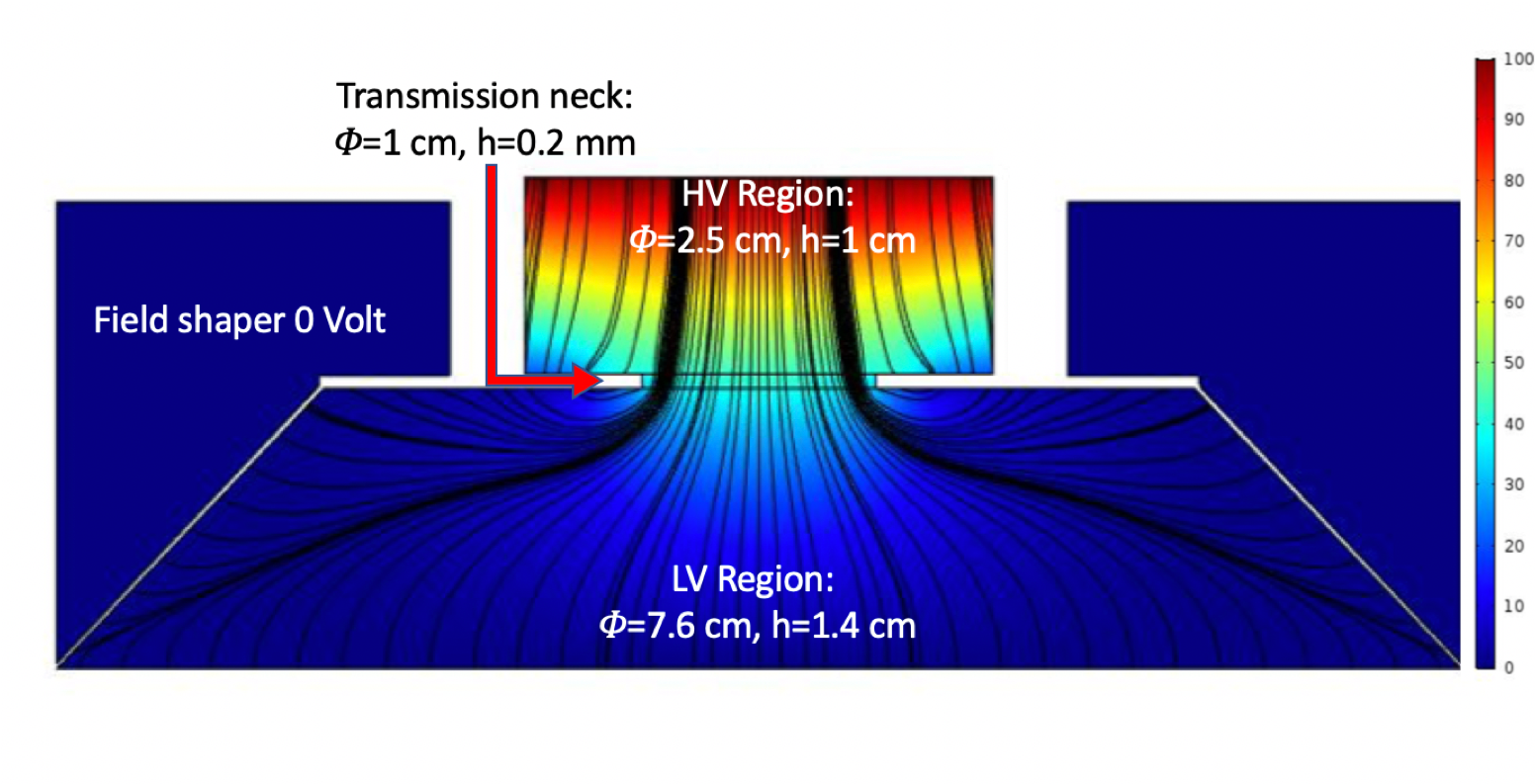}
	}
	
	\subfloat[]{%
		\includegraphics[width=\linewidth]{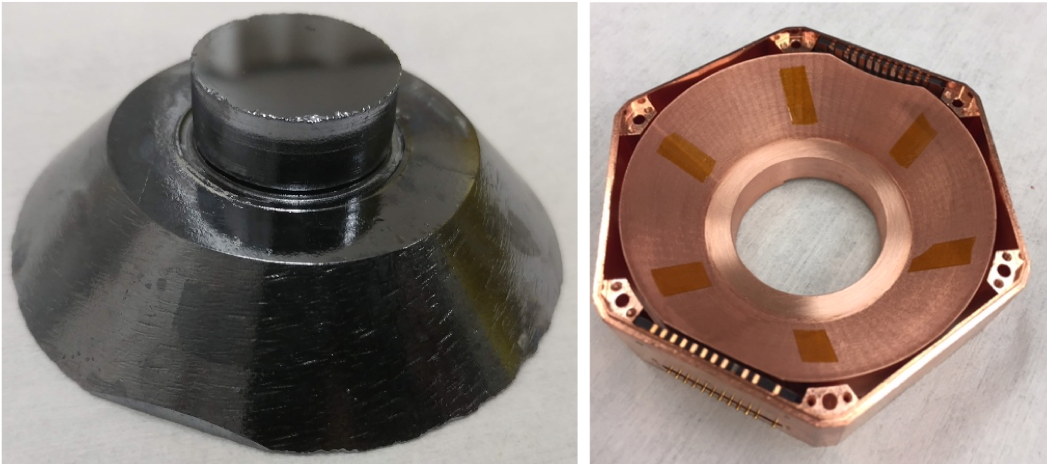}
	}
	
	\caption{(a) Simulation of electrical potential and field lines inside the detector using COMSOL Multiphysics $\textregistered$. The bias is applied from the HV top surface while the potential on the conical field shaper walls and the LV bottom surfaces are held at 0 V. The density of NTL phonons released during the carrier drift scales with the electric field along the field line paths, (b) Conical detector fabricated from a cylindrical Si crystal. The crystal is installed and secured against this copper field shaper with a conical inner wall. A 25 micron gap separates the crystal from the copper field shaper.}
	\label{fig:hybrid}
	
\end{figure}

Our hybrid detector combines the advantages of both of these SuperCDMS detector technologies, ZIP and HV,  in a monolithic crystal that is shaped such that a narrow channel divides the crystal between a high voltage (HV) region and a low voltage (LV) region as shown in Fig. \ref{fig:hybrid}. The conical LV frustrum region comprises the bulk of the detector mass and constitutes the fiducial volume for signal detection whereas the small HV cylindrical region mediates the NTL phonon generation and indirectly measures the ionization. The high voltage region provides NTL phonon-mediated indirect charge readout while the low voltage region provides recoil phonon readout. The crystal is biased from the top face while the bottom face is maintained at ground. The conical surface is also held at ground potential in order to guide the field lines through the narrow neck region in the crystal (Fig. \ref{fig:hybrid}a). The detector and field shaping electrodes were optimized to maximize the charge transmission between the two regions using COMSOL Multiphysics$\textregistered$ software.

The conical shape was fabricated by grinding a 7.6 cm by 2.5 cm cylindrical crystal at an angle of $\approx$ 45$\degree$ and core drilling the top portion on a circumference, so as to obtain the 2.5 cm diameter top region. A wire saw was used to give the final shape of the neck portion with a clearance of about $0.2$ mm. The shaped crystal and the modified Cu housing that holds the conical detector are shown in Fig. \ref{fig:hybrid}b. 

\begin{figure}[h!]
  \centering
    \includegraphics[width=0.5\linewidth]{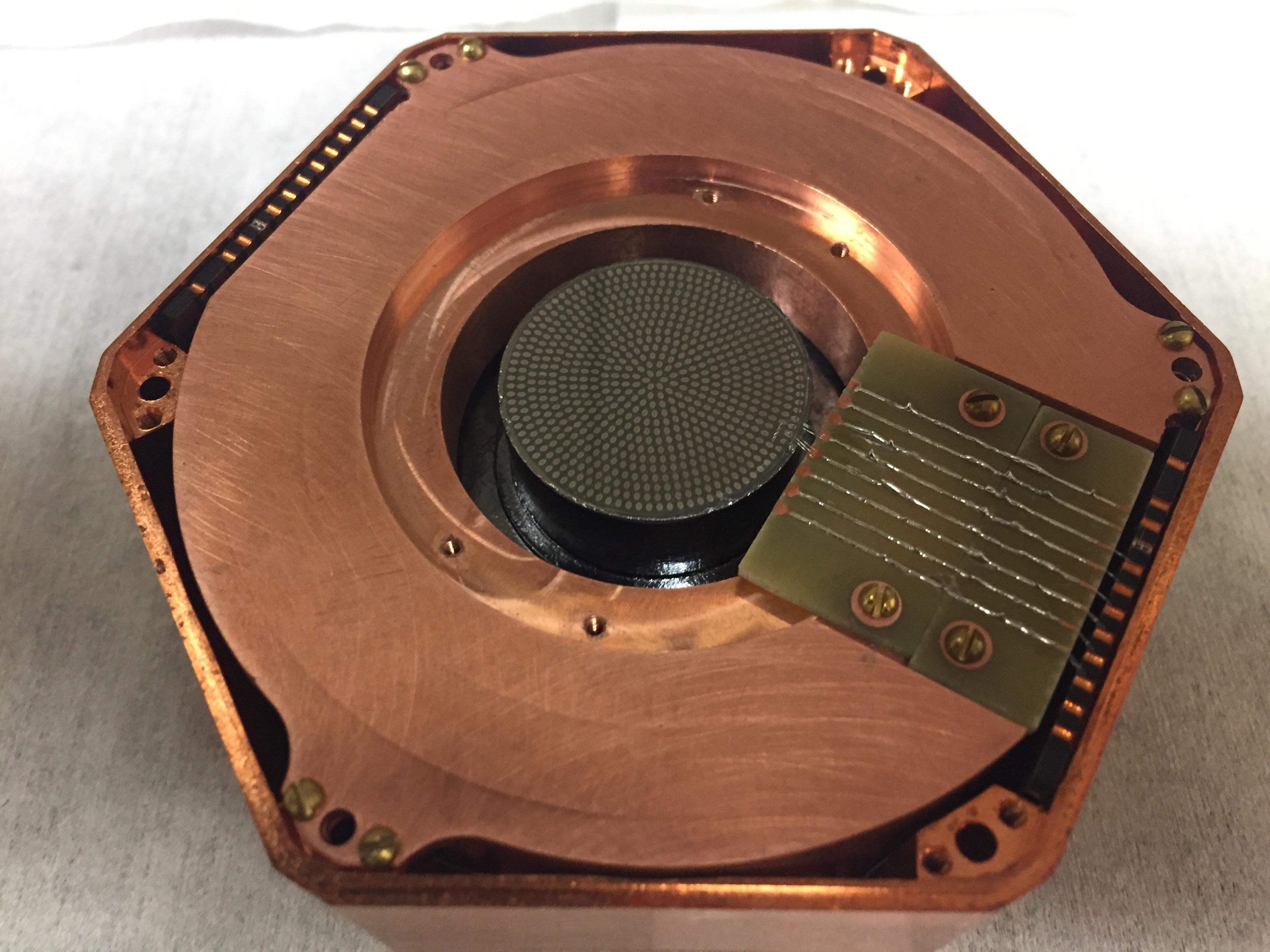}
    \includegraphics[width=0.46\linewidth]{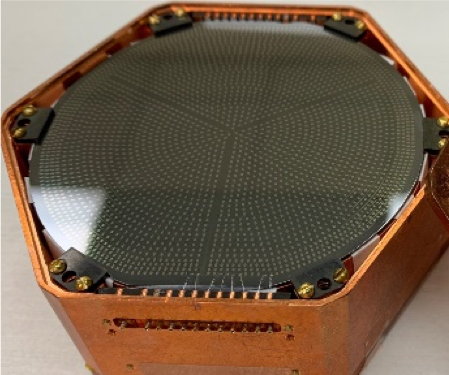}
\caption{(left) The HV face of the detector with phonon readout that is electrically floated in order to bias the detector, (right) The LV face of the detector is patterned with 4 phonon readouts which are held at ground potential.}
  \label{fig:hybridmask}
\end{figure}

CDMS style athermal phonon sensors were photolithographically patterned on the flat faces of both the HV and the LV regions of the crystal (Fig. \ref{fig:hybridmask}), in order to independently measure phonons in the HV and LV regions after each interaction. The narrow channel assisted by the field shaping electrodes surrounding the conical edges conducts charge from LV to HV region but constrains the phonon exchange between the two regions. This geometric phonon suppression, combined with full charge extraction from the LV into the HV region that then undergoes NTL gain, allows for event-by-event NR versus ER discrimination with phonon only readout and alleviates the need for noise limited high-impedance front-end charge amplifiers. 

Assuming full charge transmission efficiency and denoting the phonon leakage fraction from HV region to LV region and vice versa by $\eta_{HL}$ and $\eta_{LH}$, respectively, the phonon signal on the two sides for the events in the LV region will be given by 
\begin{equation}\label{eq:1}
P_{HV} = \alpha [(1 - \eta_{HL})E_{R}V_{HV} L/4 + \eta_{LH} E_{R}(1+V_{LV}L/4)]
\end{equation}
\begin{equation}\label{eq:2}
P_{LV} = \beta [(1 - \eta_{LH}) E_{R} (1 + V_{LV}L/4) + \eta_{HL}E_{R} V_{HV}L/4]
\end{equation}
where $\alpha$ and $\beta$ are the absolute phonon calibration factors for the top HV and the bottom LV sensors, respectively, $E_R$ is the recoil energy of the interaction, $V_{LV}$ and $V_{HV}$ are the absolute values of the effective voltage across the low voltage side and the high voltage side, respectively, and $L$ is the Lindhard factor which is normalized to 1 for ER events and has an energy dependent fractional value for NR events. 

The discrimination of the ER events which constitute the primary background can be done by a discriminating factor defined as 
\begin{equation}\label{eq:3}
Y = \frac{P_{HV}}{P_{LV}}.
\end{equation}

\begin{figure}[h!]
  \centering
    \includegraphics[width=\linewidth]{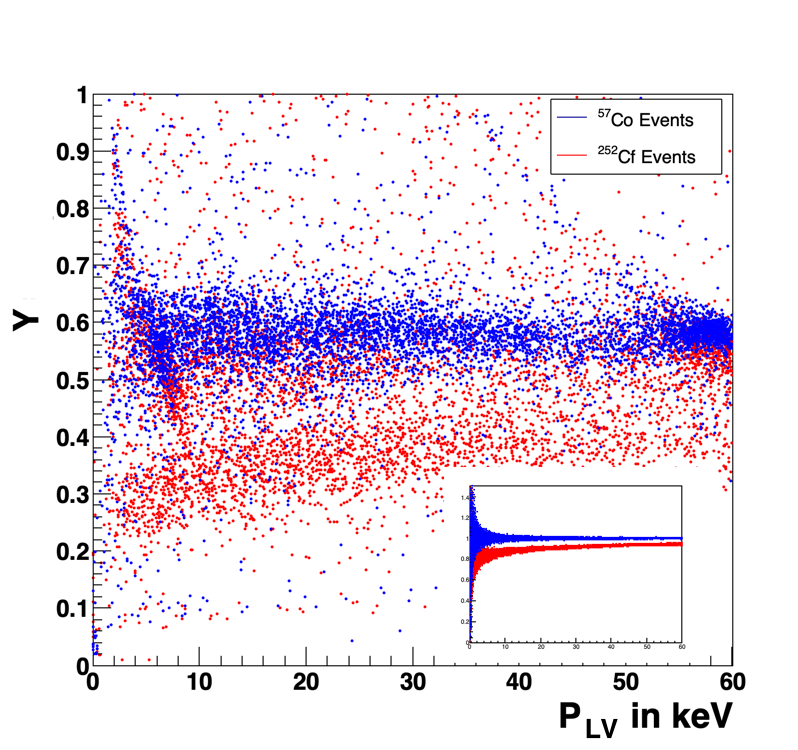}
\caption{Events from $^{57}$Co source (blue) producing ER events and a $^{252}$Cf source (red) producing both NR and ER events, without any Lindhard correction applied. (Inset) Simulated ER(blue) and NR(red) bands expected for interaction in the low voltage region based on values of $\eta_{HL}$, $\eta_{LH}$, $\alpha$ and  $\beta$ and baseline resolutions of the top and bottoms channels. NR events from $^{252}$Cf become sparse at higher energies.  
}
  \label{fig:simulation}
\end{figure}

The expected phonon signal from NR and ER events inside the detector were simulated for various bias voltages and expected phonon passage fractions from purely geometric suppression for qualitative distributions. After data collection at a particular voltage, phonon partition and the effective voltage at the interface between HV and LV regions were obtained directly from the data. Combined with the measured phonon resolutions of the HV and LV channels, updated expected bands were created, as shown in Fig. \ref{fig:simulation} (inset).


Demonstration of the detection technique relied on two key aspects: charge from the LV region could be transported to the HV region with good efficiency, and at the same time, the density of NTL phonons in the HV region do not significantly pollute the primary phonons in the LV region. Prior to the deposition of phonon sensors on the detector faces, we tested the charge transmission aspect of the detector by using a vacuum electrode ionization readout technique ~\cite{Mirabolfathi:2015pha}. With standard CDMS cold FET front-end charge amplifier, we demonstrated close to complete ionization transmission between the two regions by examining ionization pulse amplitude distribution from a collimated $^{241}$Am source as the calibration source. In addition to charge transmission, we tested the HV sustainability of the crystal for ionization leakage performance using an external power supply connected through a custom designed adaptor board with a voltage range from 0 to 400 V. No catastrophic leakage was observed up to 400 V, similar to what was observed with a similar germanium crystal fully characterized up to 400 V~\cite{Mirabolfathi:2015pha}.

To study the physics of phonon transport in this new detector geometry, tungsten superconducting ($T_c$ $\simeq$40 mK) Transition-Edge Sensors (TES) were photolithographically patterned onto the HV and LV surfaces of the Si crystal (Fig. \ref{fig:hybridmask}). The TES design was adapted from detector development work for the SuperCDMS SNOLAB experiment. The 2.5 cm HV side of the detector was designed to have one single channel with half as many TES sensors ($\approx$ 800) as a nominal phonon channel ($\approx$ 1600) in the bottom. 

The 7.6 cm LV side of the detector was divided into 4 phonon channels, with one outer veto ring and three inner channels (Fig.\ref{fig:hybridmask}). The three inner phonon channels were used for readout with SQUIDs connected to standard CDMS electronics available at the facility, while the outer phonon channel was kept at floating potential. A $^{55}$Fe source was collimated onto the HV face and provided low energy ($\approx$ 6 keV) photons, and an $^{241}$Am source was collimated onto the LV face and provided medium energy ($\approx$18-22 keV lines and a 60 keV line) photons. The penetration depth of the 6 keV photons is low ($<$100 microns), hence those photons are entirely absorbed in the HV region. On the other hand, the 60 keV photons penetrate across the 2.5 cm thick Si and interact in both LV and HV regions, albeit more in the LV region where the source is located. An external $^{57}$Co source was used to provide uniform bulk illumination to characterize the ER events in the detector. An external $^{252}$Cf source was used to provide spontaneous fission neutrons for NR calibration. 

\begin{figure}[h!]
  \centering
     \includegraphics[width=1\linewidth]{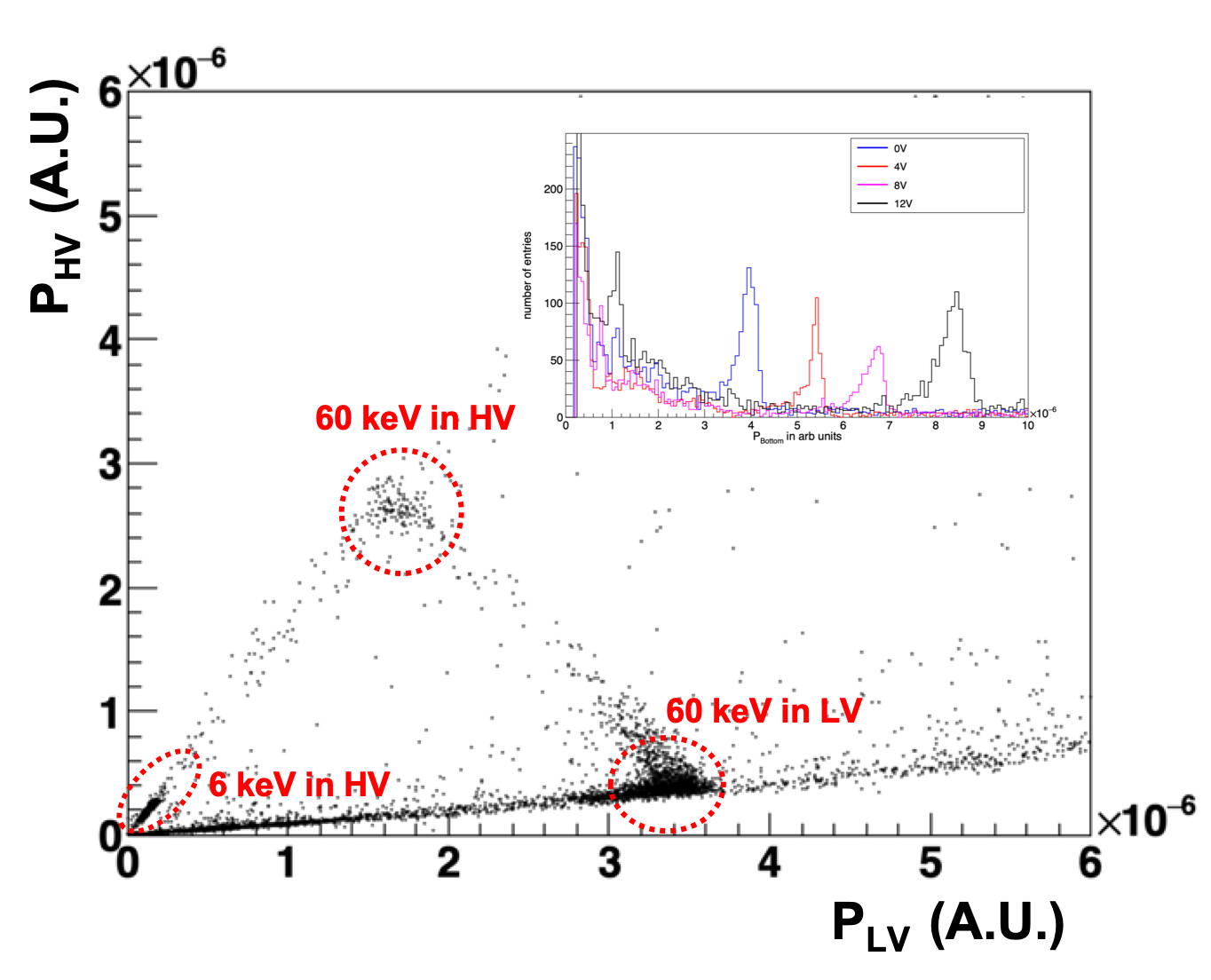}
 \caption{Distribution of phonon energy among the HV and LV phonon channels at 0 V bias that provide the key phonon transport parameters for this hybrid detector. Two distinct population of events distributed along two lines that cross the origin correspond to  events in either HV or LV regions. The points distributed along the diagonal that connects the two 60 keV populations correspond to the 60 keV photons in the transition region between the HV and the LV regions. The 6 and 60 keV populations in the HV region and the 60 keV population in the LV region are used to determine the phonon partition model parameters. (Inset) Phonon amplitude increases linearly with the applied bias as expected from the NTL gain.}
  \label{fig:noiseNTL}
\end{figure}

Events were collected by triggering on either the HV channel or the most sensitive of the LV channels. 
Optimal filtering fitting techniques developed for the SuperCDMS experiment were used to fit the phonon pulses. Good events were selected by applying various data quality cuts, such as rejecting elevated baseline noise, thermal excursions, unstable runs and electrical glitches. Various analysis cuts, including a cut on the $\chi^2$ distribution of the pulse fitting to the pulse template, were employed to select high quality events. 


\begin{figure}[h!]
\centering
   \includegraphics[width=\linewidth]{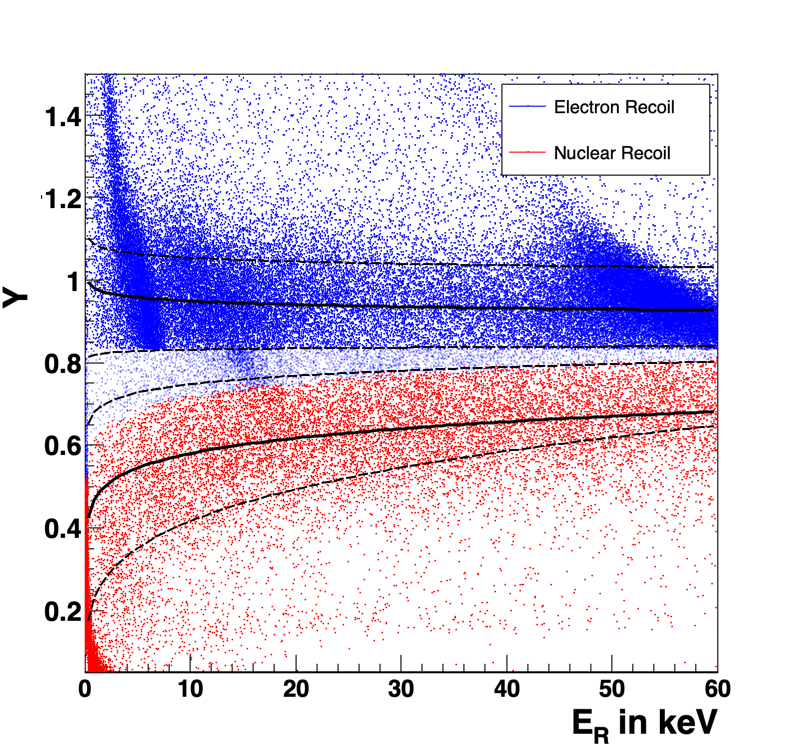}
    \caption{Events from the $^{252}$Cf calibration showing clear separation between the NR (red points) and ER (blue points). The gamma events from the $^{252}$Cf sources are visible and fall clearly in the ER band. Known populations of gamma lines at 6 keV from the $^{55}$Fe, 60 keV and a range of 14-24 keV  from $^{241}$Am sources are also clearly visible in the ER band. True recoil energy for each event falling below the ER band was subsequently recalculated (solving equations \ref{eq:1} and \ref{eq:2}) taking into account the standard silicon Lindhard factor for NR events. This leads to the artifact of a sharp cut off below the ER band, with events moving to the right in $E_R$ by the Lindhard factor ($\simeq$1/3). 
    The larger band widths compared to the simulation are due to the lack of phonon position correction.}
 \label{fig:co57Cf252}
\end{figure}

\begin{figure}[h!]
\centering
     \includegraphics[width=\linewidth]{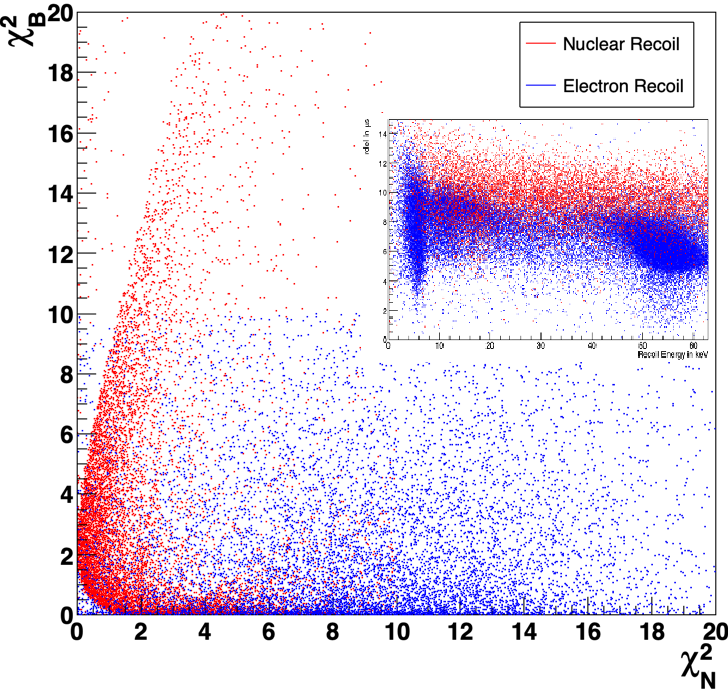}

    \caption{ER vs NR likelihood, implemented using a $\chi^2$ analysis comprising of the Y variable and the delay variable, provides a combined discriminant that can be used to define a ($\chi^2_B$-$\chi^2_N$) cut (likelihood ratio test). (Inset) Phonon delay
    provides additional discrimination between the ER and the NR events. NR events are expected to have higher delays compared to the ER events~\cite{Agnese2013_2}.}
 \label{fig:delay}
\end{figure}

We obtain $\eta_{HL}$, $\eta_{LH}$, $\alpha$ and  $\beta$ for the phonon partition model presented above by running the detector at 0 V, $\textit{i.e.}$ setting V=0 in the equations \ref{eq:1} and \ref{eq:2}, and comparing the phonon partition between the HV and LV sensors for two populations of events highlighted in Fig. \ref{fig:noiseNTL}: the 60 keV events in the HV region versus those in the LV region. Given these parameters, a run at a non-zero bias voltage sets the V$_{HV}$ and V$_{LV}$. Using these parameters, we performed a consistency check for the expected partition for the 6 keV events in the HV region and found the data consistent with the model predictions. Additionally, we found that those parameters are consistent with expectations from pure geometric suppression due to the constriction between the LV and the HV regions. By simultaneously solving the equations \ref{eq:1} and \ref{eq:2} for the measured P$_{HV}$ and P$_{LV}$, we can deduce the event recoil energy (E$_{r}$) and the Lindhard factor (L).  

We used fission neutrons from a $^{252}$Cf source to generate NR's in the detector. $^{252}$Cf neutrons have $\sim$ 1 MeV energies and  generate up to $\sim$ 100 keV NR in Si substrate. 
The NR events clearly exhibit a separated band, as expected, from the ER events, as shown in Fig. \ref{fig:simulation}. Lindhard suppression was observed for the NR events and the variation with energy was found to be consistent with what is observed in beam based calibrations with a SuperCDMS silicon detector~\cite{Agnese:2018xhe}, within uncertainties. 
Subsequently, the Lindhard function obtained for the SuperCDMS silicon detector was used as an input to define the recoil energy for each event in the NR band. 
The results from the NR calibration of this detector are shown in Fig. \ref{fig:co57Cf252}. 
The bands are wider, primarily due to lack of phonon position correction. Our detectors measure athermal phonons, thus they are sensitive to the position of the events in the detector. This position sensitivity offers fiducialization but should be corrected for energy reconstruction. 

Our future detector prototypes will benefit from fully instrumented phonon readout, whereby energy position dependencies can be corrected and removed. Nevertheless, even without full position correction, the resulting discrimination from the two bands obtained from the HV and LV regions of this hybrid detector provides a high quality discrimination between the ER and the NR events. We also checked the timing parameters associated with the ER and the NR events and found them to behave as expected from SuperCDMS detector operations, with NR events having a slower phonon propagation time compared to the ER events (Fig. \ref{fig:delay}).

We have demonstrated a new dark matter and CE$\nu$NS phonon-mediated detector technique that offers background discrimination down to unprecedented low thresholds. This discrimination improves at lower energies, thus making this new detector technology a highly desirable technology for low-mass dark matter searches and searches for coherent neutrino scattering from reactor neutrinos.  Although our method stems from the successful CDMS ionization and phonon background discrimination technology, it is purely phonon based and therefore not limited by the inherent difficulties in charge amplification instrumentation.
Our future prototype detectors will benefit from a fully instrumented athermal phonon readout, thereby better mapping out event locations within the detector volume in order to define a more accurate fiducial volume and position corrected recoil energies.

This work was fully supported by DOE grants DE-SC0017859 and DE-SC0018981. We acknowledge the contribution of the key cryogenic infrastructure (Bluefors LD400) provided by NISER, India. BM  would like to further acknowledge the support of DAE through project Research in Basic Sciences - Dark Matter and SERB-DST through J.C. Bose fellowship.


\bibliography{mybibfile}

\end{document}